\documentclass[twocolumn,footinbib,floatfix,aps,10pt,pra,reprint]{revtex4-1}
\usepackage[usenames,dvipsnames]{xcolor} 
\usepackage{amsmath,amssymb}
\usepackage{mathtools}
\usepackage{amsfonts}
\usepackage{bm}
\usepackage{bbm}
\usepackage{color}
\usepackage{latexsym}
\usepackage[english]{babel}
\usepackage{stmaryrd}
\usepackage{psfrag,graphicx}
\usepackage{subfigure}
\usepackage{natbib}
\usepackage{appendix}
\usepackage{enumitem}
\usepackage[utf8]{inputenc}
\usepackage[normalem]{ulem}
\usepackage{microtype}
\usepackage{amsthm}
\usepackage{chapterbib}
\definecolor{mygrey}{gray}{0.35}
\definecolor{myblue}{rgb}{0.2,0.2,0.8}
\definecolor{myzard}{cmyk}{0,0,0.05,0}
\definecolor{mywhite}{rgb}{1,1,1}
\definecolor{myred}{rgb}{0.9,0.1,0.}
\usepackage[colorlinks=true,citecolor=myblue,linkcolor=myblue,urlcolor=myblue]{hyperref}

\theoremstyle{definition}

\newcommand{\diag}{\operatorname{{diag}}} 
\newcommand{\ket}[1]{\vert #1 \rangle} 
\newcommand{\ketbra}[2]{\vert #1 \rangle \hspace{-2pt}\langle #2 \vert}

\renewcommand{\vec}[1]{\text{\boldmath$#1$}}

\newcommand{\eref}[1]{(\ref{#1})}

\newcommand{\Fref}[1]{Fig.\ref{#1}}

\newcommand{\llrr}[1]{\ensuremath{\left( #1\right)}}
\newcommand{\llrrq}[1]{\ensuremath{\left[ #1\right]}}

\newcommand{\expectS}[2]{\ensuremath{\left \langle #1 \right \rangle_{#2}}}

\DeclareMathOperator{\Tr}{Tr}

\begin{document}



\title{Efficient simulation of finite-temperature open quantum systems}

\author{D. Tamascelli$^{1,2}$, A. Smirne$^{1}$, J. Lim$^{1}$, S.~F. Huelga$^{1}$, and M.~B. Plenio$^{1}$ }
\affiliation{$^1$Institut f\"{u}r Theoretische Physik, Albert-Einstein-Allee 11,
Universit\"{a}t Ulm, 89069 Ulm, Germany}
\affiliation{$^2$Dipartimento di Fisica ``Aldo Pontremoli'', Universit\`a degli Studi di Milano, via Celoria 16, 20133 Milano, Italy}

\date{\today}

\begin{abstract}
Chain-mapping techniques {in combination with the time-dependent density matrix renormalization group are}
a powerful tool for the simulation of open-system quantum dynamics. {For finite-temperature environments,
however, this approach} suffers from an unfavorable algorithmic scaling with {increasing} temperature. We prove that the system dynamics under thermal environments can be non-perturbatively described by temperature-dependent system-environmental couplings with the initial environment state being in its pure vacuum state, instead of a mixed thermal state. As a consequence, as long as the initial system state is pure, the global system-environment state remains pure at all times. The {resulting} speedup {and relaxed memory
requirements of} this approach {enable the efficient} simulation of open quantum systems
interacting with highly structured environments in any temperature {range}, {with applications extending from quantum
thermodynamics to quantum effects in} {mesoscopic systems.}

\end{abstract}

\maketitle

Quantum systems are never completely isolated and the interaction with surrounding uncontrollable
degrees of freedom can modify significantly their dynamical properties. {In some cases the
environment can be assumed to be memoryless, in which case} {master equations of Lindblad form} provide an accurate
effective description of the resulting open-system dynamics \cite{car93,breuer02,gardiner04,rivas12}.
{Generally,} however, the {description} of the evolution of open quantum systems (OQSs) requires
to take into full account the environmental degrees of freedom and their interaction with the system.
{This becomes particularly important,} when the system-environment coupling is not weak, and the
environment reorganization process occurs on a time scale which is comparable to the system dynamics
{-- a situation that is ubiquitous in soft or condensed matter, nanothermodynamics} and quantum
biology \cite{huelga13,rivas14,breuer16,deVega17}. In this case, the OQS dynamics {is not accessible
to either analytical methods (apart from very few specific instances \cite{luczka90,hu92,garraway97,fisher07,smirne10,diosi14,ferialdi16}), nor effective master equation
approaches and more refined numerical techniques are thus needed.}

Over the last two decades, a variety of {numerically exact} approaches for the simulation of open
quantum systems have been proposed. These methods allowed for the {description of features that
were not accurately described by approximate methods, such as the Markov, Bloch-Redfield or perturbative
expansion techniques \cite{breuer02}.} In particular, the Time Evolving Density operator with Orthogonal
Polynomials (TEDOPA) \cite{prior10,chin10} algorithm is a certifiable method
\cite{woods15} for the nonperturbative simulation of OQS {that} has found application {for the description
of} a variety of open quantum systems \cite{prior10,prior13,chin13}. TEDOPA belongs to the class of
chain-mapping techniques \cite{hughes09,prior10,chin10, martinazzo11,woods14,ferialdi15}{, and is closely related to Lanczos tridiagonalization (see \cite{deVeg15} and references therein); these techniques}  { are}
based on a unitary mapping of the environmental modes onto a chain of harmonic oscillators with nearest-neighbor
interactions. The main advantage of this {mapping is the more local entanglement structure which results
in an} improved efficiency of density matrix renormalization group (DMRG) {methods} \cite{white92}. While TEDOPA is
very efficient {at} zero temperature{, a regime} that is hard to access by other methods such as hierarchical
equations of motion {(HEOM)} \cite{kubo89,ishizaki09,tanimura06} and path integral methods \cite{feynman48,makri92,NalbachET2010},
{its original} formulation suffers from a unfavorable scaling {when increasing the} temperature of the bosonic bath.
Because of this, other {approaches}, such as  {HEOM}, are {currently the
method of choice} in the high temperature regime.


In this work we {derive a formulation of TEDOPA for finite-temperature bosonic environments
that allows for its extension to arbitrary temperatures without loss of efficiency.} Our approach
relies on the equivalence between the reduced dynamics of an OQS interacting with a finite-temperature
bosonic environment {characterised by some} spectral density and the dynamics of the same {system}
interacting with {a zero temperature environment} and a {suitably} modified spectral density
\cite{diosi98, yu04,blasone11,deVega15}, and further exploits fundamental {properties of the theory
of orthogonal polynomials} \cite{gautschi94,chin10,woods14}.


\textit{Spectral density thermalization.---}
{Consider} a quantum system $S$ interacting with a bosonic
environment; for each environmental mode at frequency $\omega \geq 0$ the annihilation and
creation operators $a_\omega, a_\omega^\dagger$ satisfy the commutation
relations
$[a_\omega,a_{\omega'}^\dagger]=\delta_{\omega\omega'},[a_\omega,a_\omega']=[a_\omega^\dagger,
a_{\omega'}^\dagger] = 0, \forall \omega,\omega' \geq 0$. The system-environment total
Hamiltonian is defined by ($\hbar = 1$)
\begin{align}
    H_{SE} &= H_S + H_E + H_I \label{eq:totHam}\\
    H_E &= \int_0^{+\infty} d\omega \ \omega a_\omega^\dagger a_\omega; \label{eq:freeHamEnv} \,\,\,
    H_I  = A_S \otimes \int_0^{+\infty} d\omega \  O_\omega,
\end{align}
where $H_S$ is the free system {(arbitrary)} Hamiltonian
and $H_E,\ H_I$ describe, respectively, the free evolution of the environmental degrees of freedom and the {bilinear}
system-environment interaction \cite{leggett87}. In what follows we assume that { $O_\omega$ is a self-adjoint operator and, in particular, is given by:} 
\begin{equation}
    O_\omega = \sqrt{J(\omega)}X_\omega = \sqrt{J(\omega)}\llrr{a_\omega
    +a_\omega^\dagger}, \label{eq:intop}
\end{equation}
{while $A_S$ is a generic operator on the open system $S$.}
The function $J(\omega):\mathbb{R}^+\mapsto \mathbb{R}^+$ is
{defined by} the product of the interaction strength between the system and the {environmental} mode at frequency
$\omega$ and the mode density, and is usually referred to as the \emph{spectral density}
(SD) \cite{breuer02}.

{At time $t=0$, system and environment are assumed to be in a factorized state} $\rho_{SE}(0) =
\rho_S(0) \otimes \rho_E(0)$, where $\rho_S(0)$ is an arbitrary (pure or mixed) initial
state of the system, {$\rho_E(0) = \bigotimes_{\omega} \exp(-\beta \omega a_\omega^\dagger
a_\omega)/\mathcal{Z_{\omega}}$ is the thermal state of the environment at inverse temperature
$\beta = 1/k_B T$ and $\mathcal{Z_{\omega}} = \Tr_E( \exp(-\beta \omega a_\omega^\dagger a_\omega))$.}
Under these assumptions, the open system state $\rho_S(t) = \Tr_E(\rho_{SE}(t))$
{at a generic time $t$ is
{entirely} determined} by the spectral density $J(\omega)$ and the inverse temperature $\beta$
\cite{feynman63,vankampen74,breuer02,gasbarri18}. In fact,
$\rho_S(t)$ is fully determined by the {two-time
correlation function}
\begin{align}
    S(t) & = \int_0^{+\infty} d\omega \
    \expectS{O_\omega(t)O_\omega(0)}{\rho_\omega(\beta)}\label{eq:intTTCF} \\
    &= \int_0^{+\infty} d\omega\  J(\omega) \llrrq{e^{-i\omega t} (1+n_\omega(\beta))+ e^{i \omega t}
n_\omega(\beta)},
    \nonumber
\end{align}
where $O_\omega(t) = \exp(i H_E t)O_\omega \exp(-i H_E t)$ is the environmental interaction
operator evolved at time $t$ via the free Hamiltonian $H_E$ and $n_\omega(\beta) =
\expectS{a_\omega^\dagger a_\omega}{\rho_\omega(\beta)} = (\exp(\beta \omega)-1)^{-1}$.
It is then clear that given two environments with the same two-time correlation functions,
the corresponding reduced dynamics coincide \cite{car93,breuer02,tamascelli18}.

If we formally extend the integral in \eqref{eq:intTTCF} to the whole real axis and define
the anti-symmetrized spectral density $J^{\text{ext}}(\omega)=\text{sign}(\omega)J(|\omega|)$
with support on the whole real axis \cite{may04}, the two-time correlation function can be
{reexpressed in the form}
\begin{align}
    S(t) = \int_{-\infty}^{+\infty} d\omega \  \frac{J^{\text{ext}}(\omega)}{2} \llrr{1+\coth\llrr{\frac{\beta
    \omega}{2}}} e^{-i\omega t}.
    \label{eq:extTTCF}
\end{align}
It is {crucial} to note that this function can be associated with an extended bosonic environment,
with positive and negative frequencies, {governed by} $H_{E^{\text{ext}}} = \int_{-\infty}^{+\infty}
d\omega \omega a_\omega^\dagger a_\omega$, which is initially in the vacuum state (i.e., $a_\omega \ket{0}
\,\, \forall \omega \in \mathbb{R}$) and which interacts with the system via the interaction Hamiltonian
$H_{I}(\beta) = A_S \otimes \int_{-\infty}^{+\infty} d\omega \ \sqrt{ J_\beta(\omega)} X_\omega$, {and}
{that now involves} a {temperature-dependent} spectral density (T-SD)
\begin{align}
    J_\beta(\omega) = \frac{J^{\text{ext}}(\omega)}{2} \llrr{1+\coth\llrr{\frac{\beta
    \omega}{2}}}.
    \label{eq:thermalSD}
\end{align}
We conclude that the reduced dynamics in the presence of an initial thermal state of the
environment and a global Hamiltonian as in Eqs.\eqref{eq:totHam} and \eqref{eq:freeHamEnv}
is the same as {the one resulting } from an initial vacuum state of the extended
environment and {a coupling governed by the new spectral density defined in Eq.\eqref{eq:thermalSD}.
Note that, in contrast to} previous approaches \cite{diosi98, yu04,blasone11,deVega15}, we
{achieved this {equivalence} by {suitably redefining} the spectral density, which is the central object
in TEDOPA. Importantly, the relationship between the original thermal chain and the pure
state chain} with the {temperature-dependent}  spectral density can be formulated in terms of a unitary
equivalence, which, {in principle}, allows one to recover the state of the {full system-environment state
in the original picture at any time $t$} \cite{suppl}.

\textit{Thermalized TEDOPA.---}
TEDOPA \cite{prior10,chin10,tama15,kohn18} {relies on} {the theory of orthogonal
polynomials} \cite{gautschi04} to provide an analytical unitary transformation {mapping the
original star-shaped system-environment model into a one dimensional configuration \cite{chin10}.} New modes
with creation and annihilation operators $c_n^\dagger$ and $c_n$ are defined as $c_n^{(\dagger)}
= \int_{0}^{+\infty} d\omega\  U_n(\omega) a_\omega^{(\dagger)}$ using the unitary transformation
$U_n(\omega) = \sqrt{J(\omega)} p_n(\omega)$ where $J(\omega)$ is an {input (arbitrary)} SD, and $p_n(\omega),\ n=0,1,
\ldots$ are orthogonal polynomials with respect to the measure, i.e. the positive valued function,
$d\mu(\omega) = J(\omega) d\omega$ on $\mathbb{R}^+$. Thanks to the three-term recurrence relation
satisfied by the orthogonal polynomials $p_n(\omega)$, the $H_{SE}$ Hamiltonian {in Eq.}\eref{eq:totHam}
is mapped \cite{chin10} into a \textit{chain} Hamiltonian $H^C = H_S + H_I^C+ H_E^C$ with
\begin{align}
    H_I^C & = \kappa_0 A_S (c_0+c_0^\dagger)\nonumber\\
    H_E^C & = \sum_{n=0}^{+\infty} \omega_n c_n^\dagger c_n +  \sum_{n=1}^{+\infty} \kappa_n (c_n^\dagger c_{n-1} + H.c.).\label{eq:hamEnvChain}
    \end{align}
{After the unitary transformation, thus, the system interacts only with the new mode $c_0^{(\dagger)}$, and all the interactions are nearest neighbour.} The mode frequencies  $\omega_n$ and {couplings} $\kappa_n$ are related to the recurrence
coefficients for the polynomials $p_n(\omega)$ and can be computed either analytically or via
stable numerical routines \cite{gautschi94,chin10}. The crucial observation {at this point} is that{, assuming $\int_0^{+\infty} d\omega J(\omega)/\omega < \infty$, i.e. finite reorganization energy,}
the  {temperature-dependent} spectral density {in Eq.}\eref{eq:thermalSD} defines a measure $\mu_\beta(\omega)=J_\beta(\omega)d\omega$,
with support extending, by construction, over the whole real axis. {Hence, there} exists a
family of polynomials $p_{\beta,n}$ which are orthogonal with respect to $d\mu_\beta$ {and we
can} define the unitary transformation
\begin{align}
    U_{\beta,n}(\omega) &= \sqrt{J_\beta(\omega)} p_{\beta,n}(\omega) \label{eq:unitaryTemp} \\
    c_{\beta,n}^\dagger & = \int_{-\infty}^{+\infty} d\omega\  U_{\beta,n}(\omega) a_\omega^\dagger,
    \label{eq:unitTempTrans}
\end{align}
and follow the same procedure {as before}. The resulting Hamiltonian has the same form {as Eq.\eqref{eq:hamEnvChain}, with the modes} $c_n^{(\dagger)}$ replaced by $c_{\beta,n}^{(\dagger)}$
and new coefficients $\omega_{\beta,n},\kappa_{\beta,n}$ related to the polynomials $p_{\beta,n}$.
{The unitary transformations $U_n(\omega)$ and $U_{\beta,n}(\omega)$ respectively determine the
initial state of the chain: for standard TEDOPA the thermal state of the environment is mapped to the
thermal state of the chain $\rho_E^C(\beta) = \exp(-\beta H_E^C)/\mathcal{Z}_E^C$, while the vacuum
state of the extended environment is mapped to a (factorized) vacuum pure state of the chain.}

\textit{Impact on simulations.---}
As long as $\rho_S(0)$ is a pure state, the {{global} state of system and chain in the T-SD approach
remains pure for $\forall t \geq 0$.} This has a major impact on the simulation of the system
dynamics via time-dependent DMRG techniques \cite{white93,rommer97},  such as the {time-evolving-block-decimation
algorithm} (TEBD) \cite{vidal03,vidal04,zwolak04}. From now on we will refer to TEDOPA with T-SD
approach as T-TEDOPA. In order to fully appreciate the advantage provided by T-TEDOPA, here we
discuss the main features of {its scaling properties; a more detailed comparison of the complexity
of the standard and thermalized {methods} is reported {in the {SM}} \cite{suppl}.}

In order to enable computer simulations, {both} the length of the {harmonic chain
and the local dimension of the environmental oscillators must be truncated}. These truncations must be chosen
such that finite-size effects remain negligible during the simulation interval $[0,t_{\text{max}}]$.
For a chain of length $N$ and local dimension $d$, the complexity of the standard TEDOPA approach
scales as $O(N t_\text{max}(d^2 \chi)^3)$,  {where} $\chi$ {is} the \emph{bond dimension}, a TEBD parameter
that is related to the amount of {correlations} in the simulated system. {On the other hand, the
complexity for T-TEDOPA will be given by $O(N' t_\text{max}(d' \chi')^3)$, where the primed letters
emphasize that, in general, the local dimension, the chain length and the bond dimension will be
different from the standard case. Clearly, the reduced complexity of T-TEDOPA stems mainly from the
fact that only pure states are involved {in the simulation}, whereas for standard TEDOPA mixed states are needed.

{In addition, the local dimensions required to faithfully represent thermal state {of the chain} scales
unfavorably with the temperature, and, as a consequence, $d'$ can be taken significantly
smaller than $d$ \cite{suppl}. For all the dynamics taken into account here, the decrease of the
local dimension in the T-TEDOPA overcompensates by itself the increase of the chain length (we
usually set $N' \approx 2 N$ {due to an increased propagation speed in the chain}) and of
the bond dimension (we used at most $\chi' \approx \sqrt{2} \chi$) {\cite{suppl}.} }

{It is important to note that the Matrix Product Operator (MPO) representation of the chain
cannot be determined analytically {in general} and its preparation requires a {considerable additional} computational overhead.} This step is clearly not required by T-TEDOPA, since the factorized vacuum
state can be straightforwardly represented via Matrix Product States (MPS). {It is worth noting that
the approach developed in \cite{deVega15} shares some features of the T-TEDOPA. It allows to use pure instead of mixed states as well, but maps the positive and negative frequency environmental
degrees of freedom into two separate chains. This results in a locally 2-dimensional tensor network
with a consequent {considerable} increase of the simulation complexity, as discussed extensively in
\cite{suppl}. As a last, but practically relevant observation, we note that T-TEDOPA does not require
any change in the already existing and optimized TEDOPA codes, since it only needs a modification of
the chain coefficients.
\begin{figure}
    \subfigure[]{\includegraphics[width=.4 \textwidth]{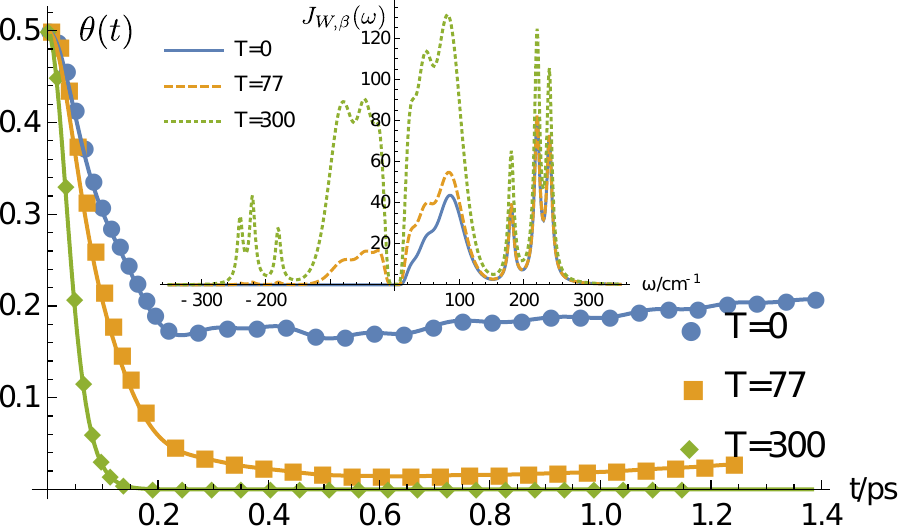}}\\
    \vspace{-0.5cm}
    \subfigure[]{\includegraphics[width=.4 \textwidth]{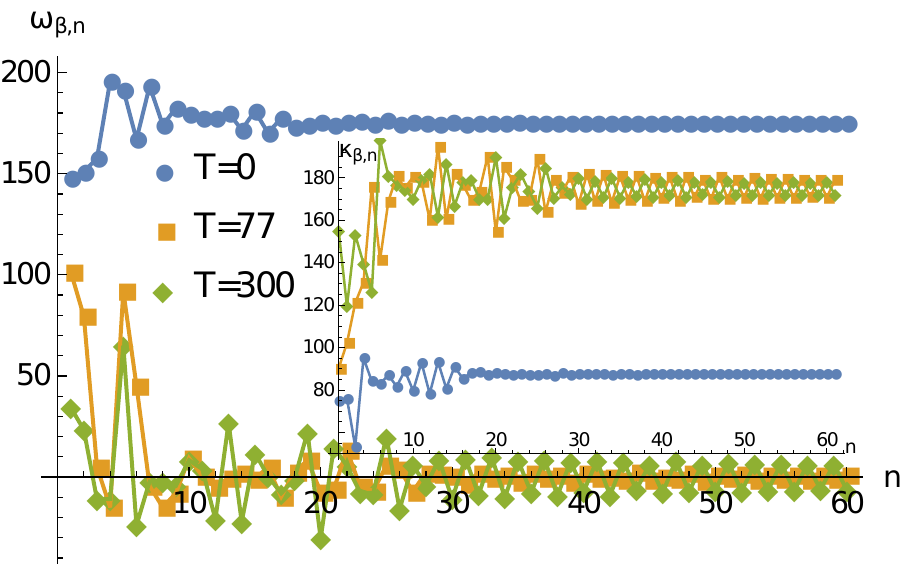}}
    \caption{\label{fig:figure1}
    Coherence dynamics $\theta(t)$ for a TLS subject to pure dephasing induced by a reservoir {modelled by} $J_W(\omega)$  at $T=0,77,300\,{\rm K}$. Markers
  represent T-TEDOPA results; analytic results,  defined as in the text, are shown as solid lines. In the inset,
   the corresponding T-SD $J_{W,\beta}(\omega)$. (b) Chain coefficients $\omega_{\beta,n}$ and $\kappa_{\beta,n}$ (inset) corresponding to $J_{W,\beta}$ for $T=0,77,300\,{\rm K}$.
    %
%
    }
\end{figure}
\begin{figure}[h]
	\subfigure[]{\includegraphics[width=.4 \textwidth]{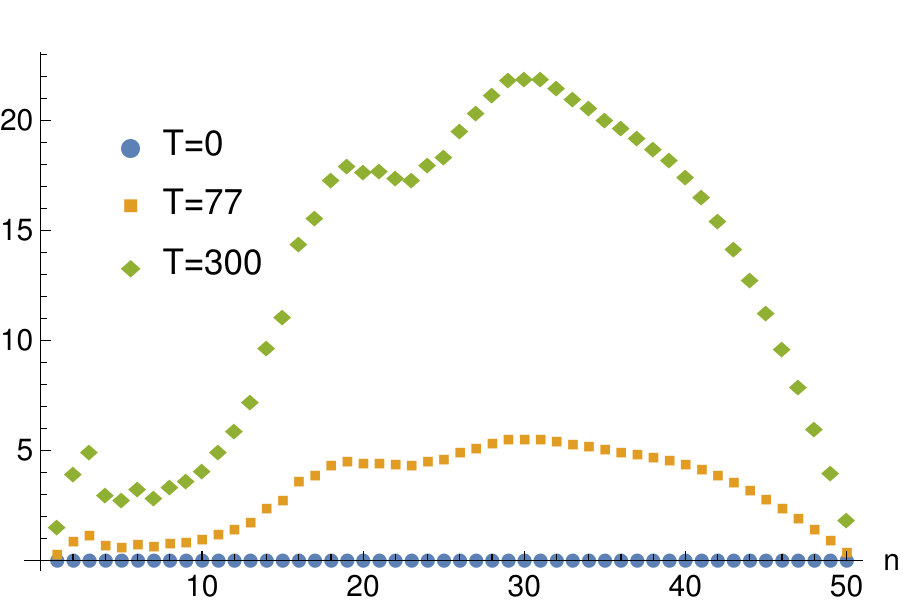}}\\
	\vspace{-0.5cm}
    \subfigure[]{\includegraphics[width=.4 \textwidth]{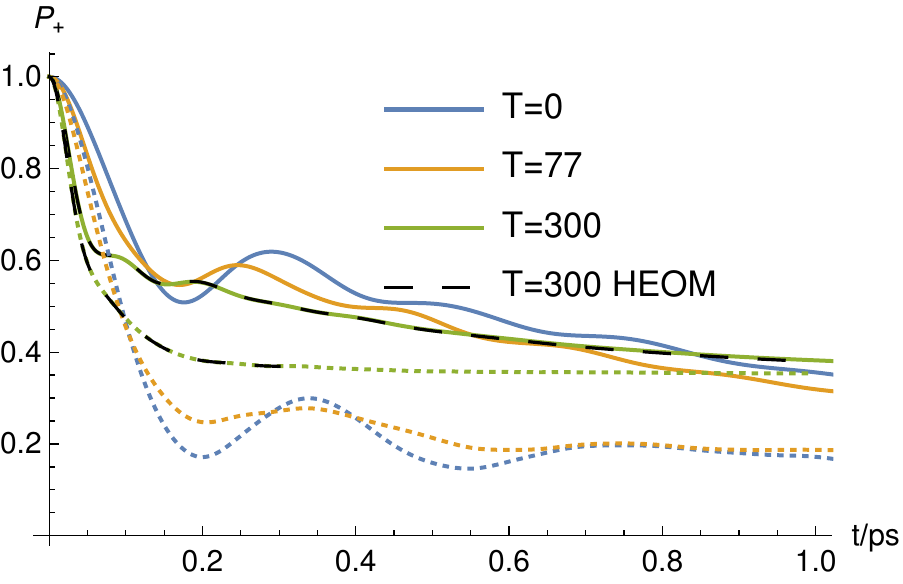}}
    \caption{\label{fig:figure2}(a)  {TEDOPA chain initialization.} Average occupation number
    of the oscillators of a chain obtained by the standard mapping of $J_W(\omega)$;
    {the values have been obtained via the procedure described in \cite{suppl}}. (b) Simulation results for a model dimeric system.
    The
    expectation value of $P_+ = \ketbra{+_D}{+_D}$ as a function of time at different temperatures shows the dynamical effect of the environmental noise on the lifetime of coherent superpositions of (electronic)
    quantum states. {Dotted}
lines correspond to a structured spectral form $J_W(\omega)$, while {solid} lines correspond to  $J_W'(\omega)$.
 { HEOM \cite{kubo89, ishizaki09,tanimura06} results are shown (dashed black lines) for comparison at $300\,{\rm K}$ (see SM {\cite{suppl}} for more details).}}
\end{figure}

\textit{Case study{.}---}
{In order to illustrate the main features of T-TEDOPA, we present two examples where} we consider
environments with a structured SD $J_{W}(\omega)$, consisting of a broad background plus three
Lorenzian peaks. 
 {This type of spectral density is characteristic of pigment-protein complexes, where electrically coupled pigments are subject to the structured environment provided by intra-pigment and protein vibrations \cite{Ulrich15,Ulrich11}.}

{The accuracy of the results provided by T-TEDOPA is clearly apparent when comparing the simulation results with a solvable model.
Consider a two level system (TLS) subject to a pure dephasing dynamics. The environment and interaction Hamiltonians are defined as in
{{E}qs.}(2) and (3) with  $A_S = (1+\sigma_z)/2$ and  {$J(\omega)=J_{W}(\omega)$}.
The  {T-SD in Eq.}\eref{eq:thermalSD} at $T=0,77$ and $300\,{\rm K}$ are shown
in the inset of \Fref{fig:figure1}(a), while its full definition is provided in \cite{suppl}. {We imposed a hard
cut-off $\omega_c=350${$\,{\rm cm}^{-1}$} {such that} $\int_{\omega_c}^\infty d\omega J_W(\omega)/\omega$ becomes
negligible ($<10^{-4}{\rm cm}^{-1}$)}.
Assume that the initial state of the TLS is a coherent superposition of the form $\ket{+} = (\ket{0}
+ \ket{1})/\sqrt{2}$. In an interaction picture, the system's coherence is given by $\theta(t) = \exp(-\gamma(t))/2$, 
with $\gamma(t) = \int_0^{\omega_c} d\omega J_{W}(\omega) \coth\llrr{{\omega}/{2 k_B T}}\llrrq{(1
- \cos \omega t)/{\omega^2}}$~\cite{breuer02} where $\gamma(t)$ is often referred to as the decoherence function.}

As clearly shown in \Fref{fig:figure1}(a), T-TEDOPA accurately reproduces
the behavior of the coherence for $t<1.4\,\text{ps}$, with maximum error $<10^{-4}$. As shown in
\Fref{fig:figure1}(b), the T-TEDOPA chain coefficients depend, as expected, on the temperature {$T$}.
In particular, we observe that the coupling $\kappa_{\beta,0}$ between the system and the first
oscillator in the chain increases with {$T$}. {For any  assigned SD $J(\omega)$,  we obtain} $\kappa_{\beta,0}
= ||J_\beta||_1 = \sqrt{\int_{-\omega_c}^{\omega_c} d\omega J_\beta(\omega)}$, which is a non-decreasing
function of $T$. Moreover, {the behaviour of the chain parameters {$\kappa_{\beta,n}$} and $\omega_{\beta,n}$ as
functions of $n$ becomes more and more jagged as {$T$} increases, inducing an effective detuning
between nearest neighboring sites in the initial part of the chain. This configuration 
leads to non-negligible back-scattering of an excitation located initially
at the first site of the chain \cite{suppl}. {A} systematic analysis of these processes, which underpin the non-Markovian part of the dynamics,
and their non-trivial temperature dependence will be the subject of a future work. Here we simply} point out that
{this configuration results in the first site{s} of the T-TEDOPA chain having a higher occupation number}. This allows a gradual decrease {in} the local dimension{s}
{$d_n'$} of the  {$n=1,\ldots,N $,} {which significantly reduces the} simulation
complexity. For example, for the simulation at $T=300\,$K, the dimension $d_n'=d_\text{max}'
-n (d_\text{max}' -2)/N$ with $d_\text{max}'=12$ ($\chi=50$) led to converged results. We notice,
moreover, that the chain coefficients $\omega_{\beta,n}$ and  $\kappa_{\beta,n}$ tend to converge
for large $n$ to the expected asymptotic values \cite{gautschi04,woods14}: if $[a(\beta),b(\beta)]$
is the support of $J_\beta(\omega)$, then $\omega_{\beta,n} \stackrel{n\to \infty}{\longrightarrow}
(a(\beta)+b(\beta))/2$ whereas $\kappa_{\beta,n}^2 \stackrel{n\to \infty} {\longrightarrow}
(b(\beta)-a(\beta))^2/16$. Since the support is $[0,\omega_c]$ at $T=0$ and $[-\omega_c,\omega_c]$
at $T>0$, this means that at finite temperature T-TEDOPA will in general require longer chains than
standard TEDOPA. This increase in length, however, {leads to a constant factor increase in the}
T-TEDOPA complexity and is {significantly overcompensated} by the possibility of starting from
the vacuum state. Indeed, as mentioned before, the local dimension of the standard TEDOPA scales
unfavorably with the temperature, as we exemplify in \Fref{fig:figure2}(a) where we show the average
occupation number of the chain {consisting of} $N=50$ {oscillators}. It is clear that the
minimal local dimension of the oscillator chain must be chosen  much larger than the average occupation
number to allow for an accurate representation of the chain thermal state. It   is not surprising that
the sole preparation of the chain thermal state at $T=77\,{\rm K}$ required one week of computation for the
choice $d=8$ (16 Intel Xeon E5-2630v3 cores), while the $T=300\,{\rm K}$ T-TEDOPA
simulation (\Fref{fig:figure1}(a)) required only 8 hours using the same  cores  \cite{suppl}.

{As a second example, we discuss} the simulation of {a form of the water-soluble chlorophyll-protein (WSCP)
homodimer, a model system for the study of pigment-protein interactions and for which there exists a rather complete experimental characterization, both structurally and in terms of its linear and   {nonlinear optical responses} \cite{renger11,donatas14}.  We model the WSCP dimer as }} two identical TLSs with interaction Hamiltonian $H_{S} = H_D
= \lambda \sigma_+^L \sigma_-^R+ \text{H.c.}$, where $\lambda=69\,\text{cm}^{-1}$ is the cross coupling
term and $\sigma_{\pm}^{L,R}$ are the spin raising and lowering operators $(\sigma_x^{L,R} \pm i
\sigma_y^{R,L})/2$ on the left ($L$) and right ($R$) TLS.
 When restricted to the single excitation
subspace, $H_D$ admits the eigenvalues $\pm 69\,\text{cm}^{-1}$ with corresponding eigenstates $\ket{\pm_D}$.
Each TLS interacts with a local harmonic bath. The two baths are independent but described by the same
spectral density $J_W(\omega)$ used so far. The interaction Hamiltonian is $H_I = H_I^{L}+H_I^{R}$
with $H_I^{L(R)}$ defined as in {Eq.}(2)
with {$A_S^{L(R)} = (1+\sigma_z^{L(R)})/2$}. Since the overall Hamiltonian
{conserves the excitation number, the evolved state} belongs to the space spanned by
$\ket{\pm_D}$. {Fig.}\ref{fig:figure2}(b) shows the evolution of the projection $P_+=\ketbra{+_D}{+_D}$
as a function of time, when the system starts from {$\rho_S(0)=\ketbra{+_D}{+_D}$}, for two different spectral
densities, namely the {full} spectral density $J_W(\omega)$ and $J_W'(\omega)$ {where only the background is considered}.
The simulation at $300\,$K required $d_\text{max}'=20$, $\chi=180$. {A detailed discussion of the influence of the Lorentzian contribution
to the reduced {system} dynamics  and the comparison with actual experiments is beyond the scope of this work, but our results
already show the capability of the method to make predictions across the whole temperature range and for {highly structured} spectral densities.}

\textit{Conclusion and outlook.---}
In this work we have presented a {new method, T-TEDOPA, for the efficient, accurate and certifiable
simulation of open  quantum {system} dynamics at arbitrary temperatures. The central insight was a suitable
redefinition of the environmental spectral density which allowed for the use of a zero temperature environment
in place of a finite temperature environment without affecting the system dynamics. This {allows
for using MPS in place of MPO for the description of the harmonic chain of environmental oscillators. As a consequence,
we obtain a significant reduction in the scaling of the algorithmic complexity as compared to state-of-the-art chain
mapping techniques and orders of magnitude reductions in computation time.} 
By construction,
 T-TEDOPA can be {implemented} as a plug-in procedure by the already
existing and highly optimized TEDOPA codes, which can now be used to efficiently simulate open quantum
system dynamics {across the entire temperature range.}

{Our approach is particularly relevant whenever one wants to provide a quantitative
description of open-system dynamics in the presence of structured and non-perturbative
environments, such as those commonly encountered in quantum biology \cite{huelga13},
nanoscale thermodynamics \cite{VinjanampathyA16} or condensed-matter systems \cite{leggett87}, as well
as situations where the effect of environmental noise has to be identified accurately
to discriminate it from possible fundamental decoherence in high-precision tests of the
quantum superposition principle \cite{bassi13,arndt14}, {or be exploited as building block in other methods, such as the Transfer Tensor scheme \cite{cerrillo13,rosenbach16}.} {Future research will be devoted to the extension of the T-TEDOPA method to more general types of system-bath interactions.}
\\

We thank Felipe Caycedo-Soler and Ferdinand Tschirsich for many useful discussions;
we  acknowledge  support by  the ERC Synergy grant BioQ, the QuantERA project NanoSpin, the EU projects AsteriQS and HYPERDIAMOND, the BMBF projects DiaPol, the John Templeton Foundation and the CINECA-LISA project TEDDI.

\newpage

\title{Supplemental Material to ``Efficient simulation of finite-temperature open quantum systems''}

\author{D. Tamascelli$^{1,2}$, A. Smirne$^{1}$, J. Lim$^{1}$ S.~F. Huelga$^{1}$, and M.~B. Plenio$^{1}$ }
\affiliation{$^1$Institut f\"{u}r Theoretische Physik, Albert-Einstein-Allee 11,
Universit\"{a}t Ulm, 89069 Ulm, Germany}
\affiliation{$^2$Dipartimento di Fisica ``Aldo Pontremoli'', Universit\`a degli Studi di Milano, via Celoria 16, 20133 Milano, Italy}

\date{\today}

\maketitle

\section*{Supplemental Material}
\subsection{Chain occupation number}\label{sec:d}
At finite temperature, TEDOPA requires {the determination of the thermal state of the
chain of environment oscillators in MPO representation. {This thermal state is determined as the fixed
point of the evolution in imaginary time under the environment Hamiltonian} \cite{zwolak04}.
For this time consuming procedure to be accurate, the local} dimension of each oscillator
in the chain must be chosen sufficiently large to allow for an accurate representation of
the thermal state. Here, we describe a general procedure to determine the chain occupation
number, which in Sect.\ref{sec:vs} will also allow us to give an estimate of the local
dimension that must be considered at the different temperatures.

Given the chain coefficients $\omega_n, \kappa_n$ associated by {the} chain mapping to an assigned
spectral density $J(\omega)$,  and the number of chain sites $N$, chosen as not to have finite size
effects, a lower bound on the local dimension of the oscillators can be provided by the following
procedure. The chain Hamiltonian (see eq. (7) in the main text)
\begin{equation}
H_E^C  = \sum_{n=0}^{N-1} \omega_n c_n^\dagger c_n +  \sum_{j=1}^{N-2} \kappa_n (c_n
c_{n+1}^\dagger + H.c.)\label{eq:hamEnvChainTrunc},
\end{equation}
can be rewritten as
\begin{equation}
    H_E^C = \vec{c}^\dagger A \vec{c}
    \label{eq:hchainMatr}
\end{equation}
with $\vec{c} = \llrr{c_0,c_1,\ldots,c_{N-1}}^T$
\begin{equation}
    A=\llrr{
        \begin{array}{c c c c c}
        \omega_0 & \kappa_1 & 0 &\dots & 0 \\
        \kappa_1 & \omega_1 & \kappa_2 & \dots &0 \\
        \dots& \dots & \dots & \dots & \dots \\
        0 & 0& 0 & \kappa_{N-2} & \omega_{N-1}
    \end{array}
}.
    \label{eq:Amatrix}
\end{equation}
The three-diagonal real matrix $A$ can be diagonalized $A=U^T D U$ with $D=\diag\llrr{\omega_0',
\omega_1',\ldots,\omega_N'}$ and $U$ {being} a unitary operator $U^T = U^{-1}$. {Stated otherwise,}
the chain admits $N$ normal modes with frequency $\omega_n',\ n=1,2,\ldots,N$  and creation/annihilation
operators defined as linear combinations of of the operators $c_n, c_n^\dagger$ by
\begin{align}
    \vec{b} &= U \vec{c}, \\
    \vec{b}^\dagger &= \vec{c}^\dagger U^T.
    \label{eq:transFiniteChain}
\end{align}
By linearity, the average occupation number of the $n$-th chain oscillator $\langle c_n^\dagger c_n
\rangle_\beta$ at inverse temperature $\beta$ can be determined from the average occupation number
of the chain normal modes $\langle b_n^\dagger b_n \rangle_\beta = 1/(e^{\beta
\omega_n}-1)$ of the normal modes $\omega_n'$ through
\begin{equation}
    \langle c_n^\dagger c_n \rangle_\beta = \sum_{k=0}^{N-1} \llrr{U_{k,n}}^2 \langle b_k^\dagger
    b_k
    \rangle_\beta,
    \label{eq:linearRelation}
\end{equation}
where $U_{k,n}$ is the element in the $k$-th row and $n$-th column of $U$.
Since the thermal state is Gaussian it is in principle possible to determine its state, and the
expectation of any observable on the chain
analytically. In particular, it would be possible to determine the average occupation of each level
of each oscillator. However, by simply using the convexity of expectation {values}, it is possible to claim
that if $\langle c_n^\dagger c_n \rangle_\beta = j$ then at least the {lowest $\lceil j \rceil$
levels} are
occupied. This suffices to provide {an estimate for the scaling} of the required local dimension of the chain
oscillators with the temperature, {see also Sect.\ref{sec:vs}}.

\subsection{{Length of the chain}}\label{sec:N}
{For a predetermined simulation time $t_\text{max}$, we need to determine the required chain
length $N$ to prevent reflections off the end of the chain to influence the system dynamics. Since
the coefficients derive from orthogonal polynomials with respect to a temperature dependent measure
$d\mu_{\beta}(\omega) = J_{\beta}(\omega) d\omega$, the value of $N$ depends on both $t_{max}$ and
the temperature. As already shown in Fig.1b of the main text, for large $N$ the chain
coefficients $\omega_{\beta,n}$ and  $\kappa_{\beta,n}$ converge towards asymptotic values that
depend only on the support of $J_\beta(\omega)$. For large $N$, in the quasi-homogeneous region of
the chain, an excitation propagates at a speed proportional to $\kappa_{\beta,\infty} \stackrel{\text{def}}{=}  \lim_{n \to \infty} \kappa_{\beta,n}$. As $\kappa_{\beta,\infty}$ tends to grow with temperature,
this explains the need for longer chains for T-TEDOPA at finite temperatures as compared to standard
TEDOPA. A better estimate of the required chain length for assigned $J_\beta(\omega)$ and $t_\text{max}$
can be obtained from a heuristic technique that turns out to be quite reliable.}

The propagation of an excitation injected by the system into the chain can be studied by a quantum
walk like approach. We consider an initial state where a single excitation is located at the first
site of the chain {of environmental oscillators and remove the system-environment coupling.} Since
the chain Hamiltonian conserves the number of excitations, the evolution of the chain {is confined
to the} single excitation sector. We can therefore use the Hamiltonian
\begin{align}
    H_{qw} = \sum_{n=0}^{M} \omega_{\beta,n} \ketbra{n}{n} + \sum_{n=1}^{M-1} \kappa_{\beta,n}(\ketbra{n+1}{n}+
    \text{H.c.}),
    \label{eq:qwHam}
\end{align}
where $\ket{n}$ indicates a chain with the excitation located at the $n$-th TLS and $M$. The evolved state
$\ket{\psi_{qw}(t)} = \sum_{n=0}^M \alpha_n(t) \ket{n}$ can be easily computed by solving a linear system
of $M$ coupled equations. The optimal length $N$ can {then be} estimated by direct inspection of the coefficients $|\alpha_n(t)|^2$ for different values of $M$. {The optimal value corresponds to the smallest
$N$ such that the excitation after reflection off the end of the chain has not reached the first site with
an appreciable probability in the time interval $[0,t_\text{max}]$.} There is a good agreement between the
position of the propagation front provided by such an approach and the actual propagation in the thermalized
chains at all the considered temperatures, as exemplified by  Fig.1 (a) and (b).
\begin{figure}
    \includegraphics[width=.49 \columnwidth]{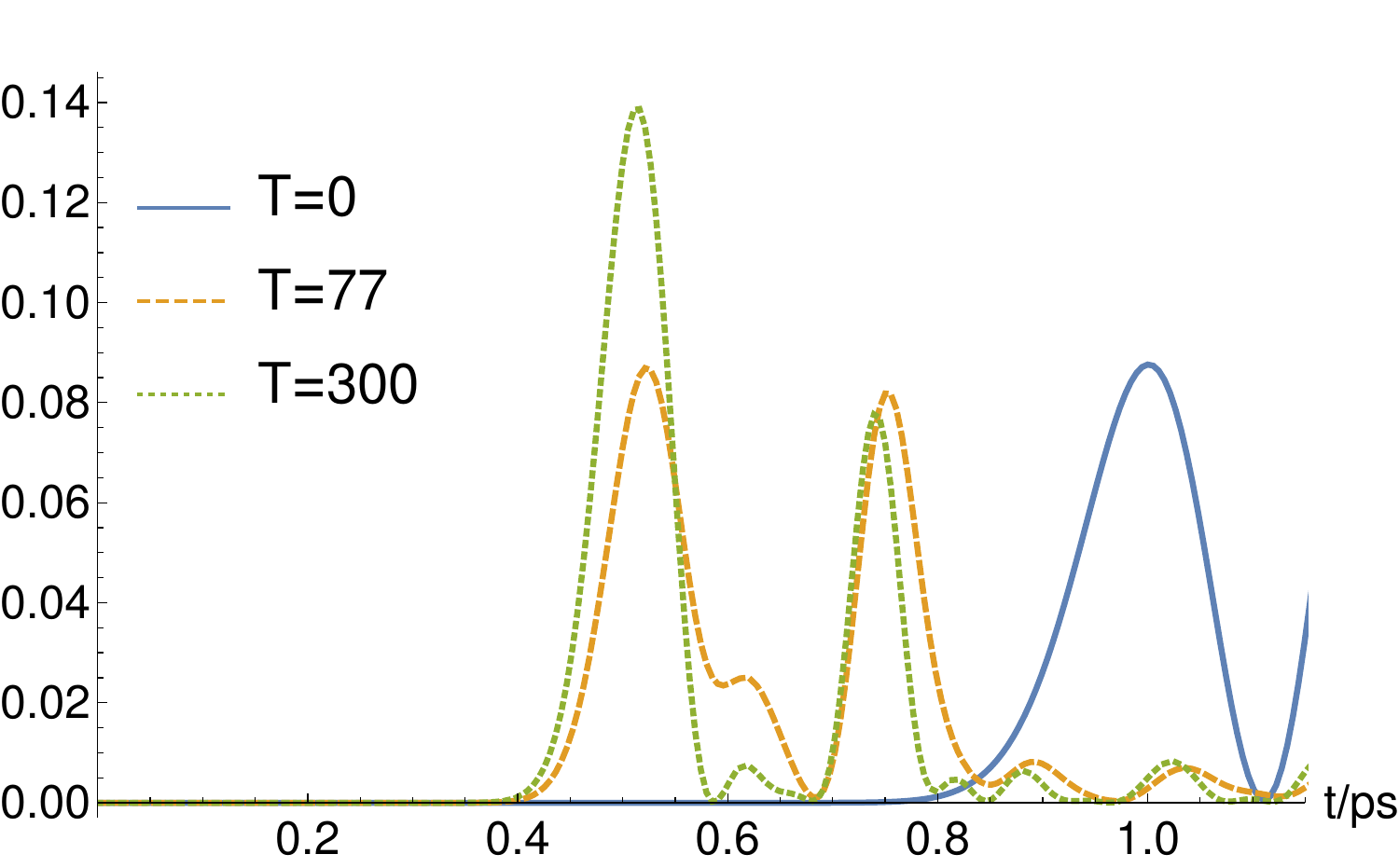} \includegraphics[width=0.22 \textwidth]{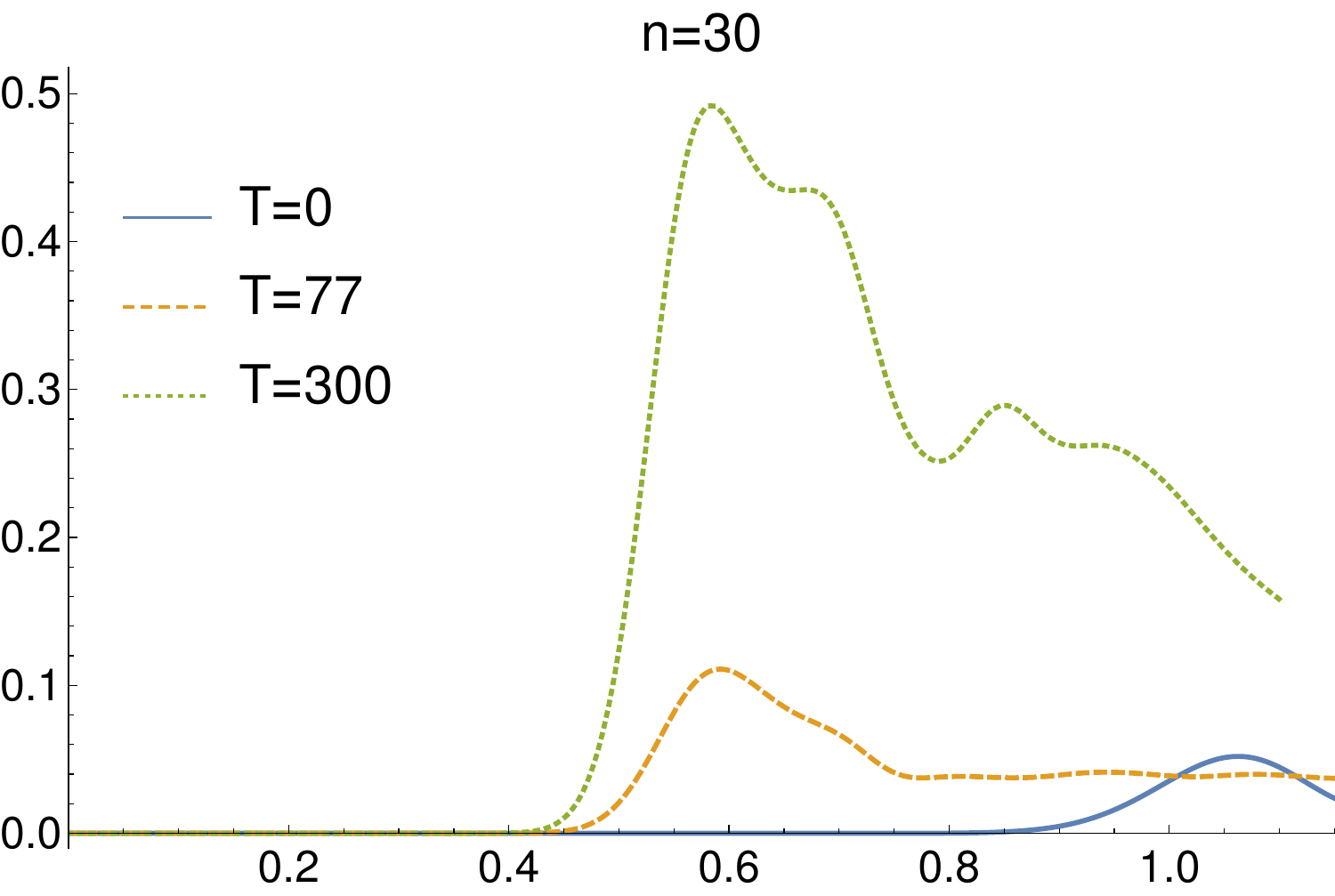}
    \caption{\label{fig:occupQW}(a) The probability $|\alpha_{30}(t)|^2$ of finding the single excitation,
    initially located in the first site $\ket{\psi_0} = \sum_{n=1}^{60} \delta_{n,1}$ of an $N=60$
    sites chain governed by the Hamiltonian (16) of the main text. (b) The actual average occupation number, as a function of time, of the 30-th chain oscillator at different temperatures.}
\end{figure}

\subsection{Excitation dynamics of the first chain sites}
Another fundamental feature of T-TEDOPA is the accumulation of excitation in the {oscillators that are
closest to the system}. This effect is {monotonous in} the temperature, as shown in \Fref{fig:occupationN}.
This can be {explained qualitatively} by looking at the couplings and the energies in this region. First
of all, as already {remarked in the main text}, the system-chain coupling is a {monotonous} function
of $T$ (see Fig.1(b) of the main text). This {implies} that, at least in the {early stages
of the evolution}, {with increasing temperature, more excitations are} created in the first {few}
chain sites by the interaction with the system. Moreover, in the same region the site energies and couplings
{exhibit significant disorder that is increasing with temperature. Therefore, excitations created by the
interaction with the system cannot propagate ballistically in this part of the chain and get partially localized,
as can be observed in}  \Fref{fig:occupationN}. This behavior {is fundamental for a chain initialized in
its vacuum state to achieve, at the level of the system-dynamics, the same dynamics as} a thermalized chain.
{We will examine this aspect of the dynamics in more detail in a forthcoming work.}
\begin{figure*}
\subfigure[]{\includegraphics[width=0.3\textwidth]{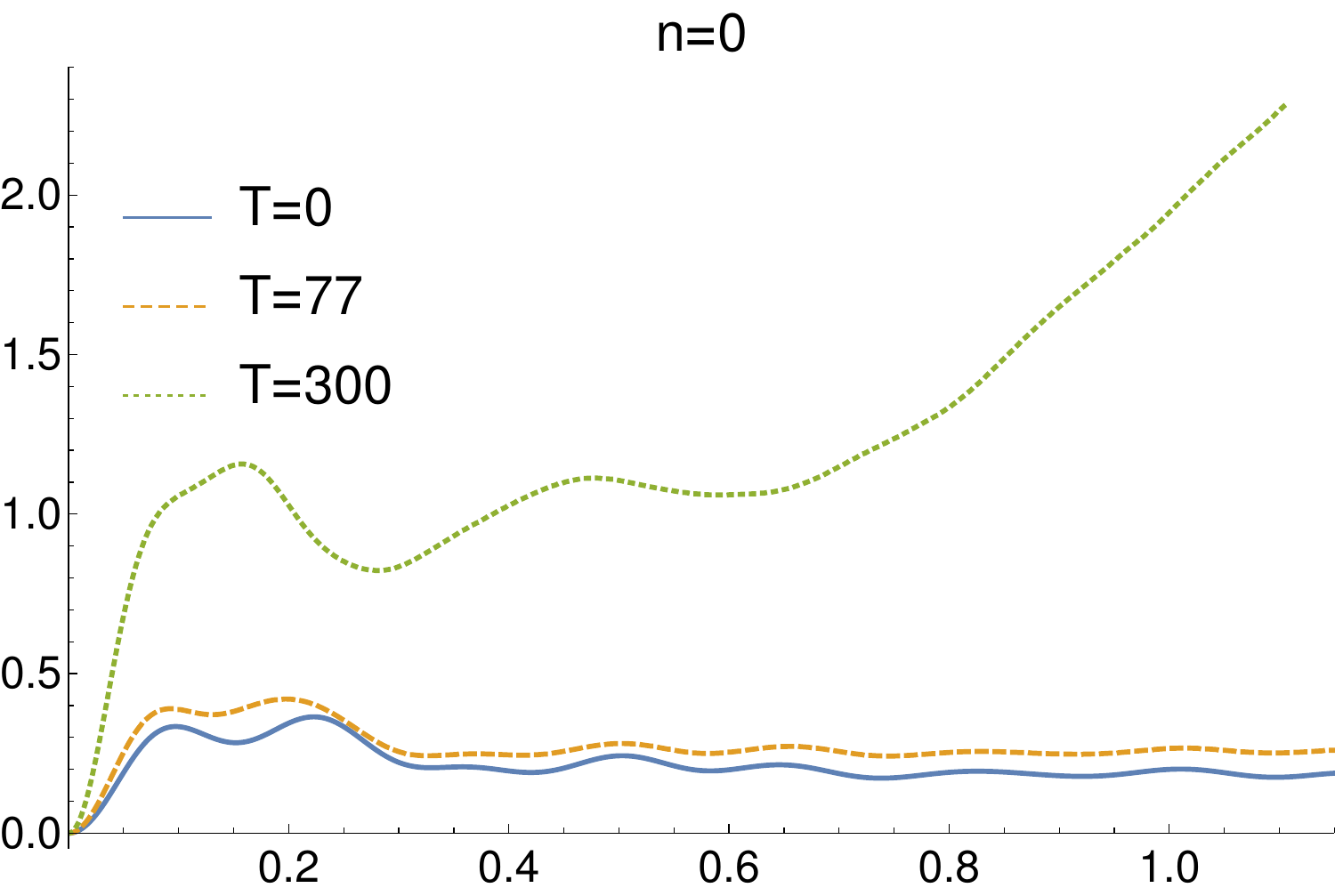}}
\subfigure[]{\includegraphics[width=0.3 \textwidth]{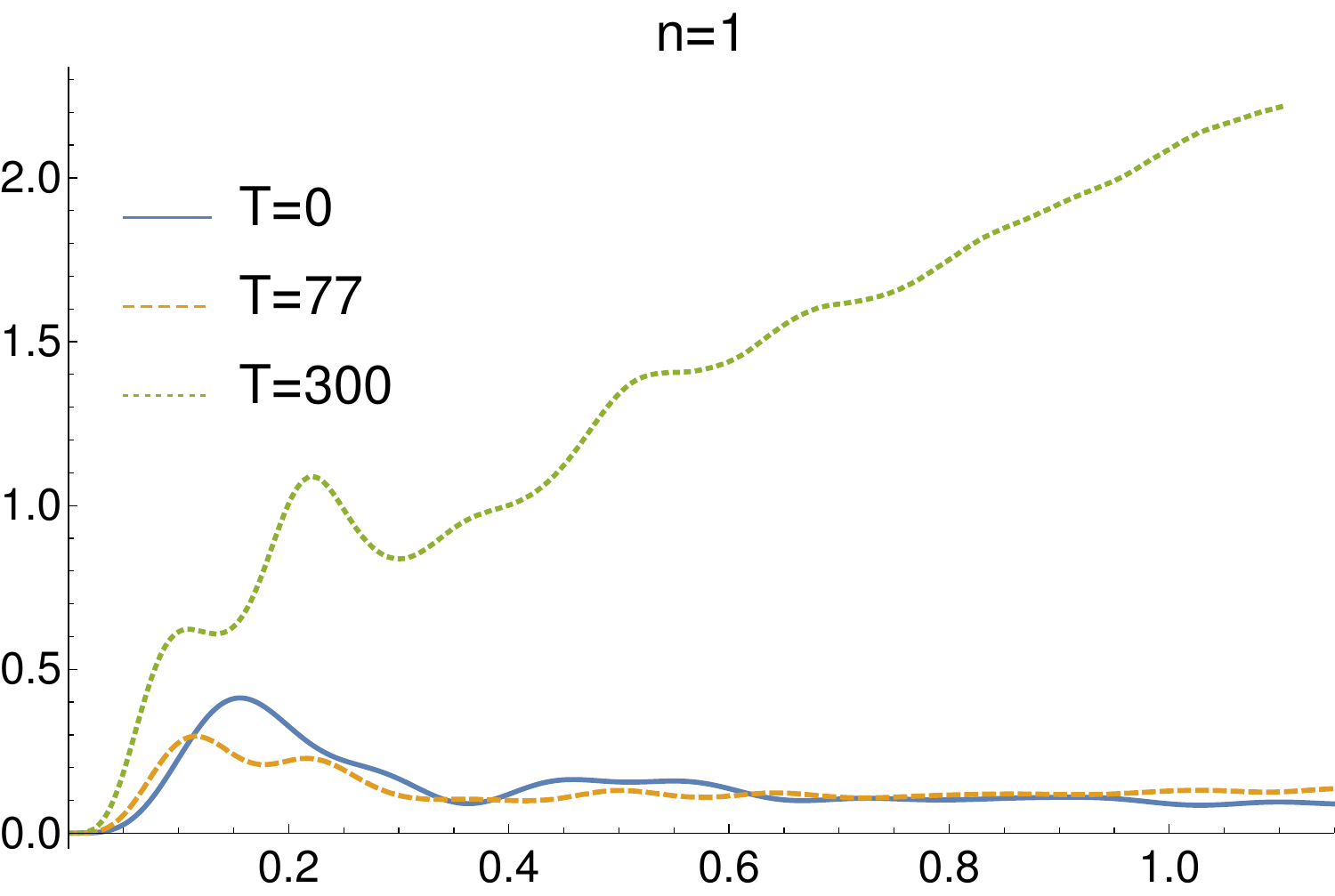}}
    \subfigure[]{\includegraphics[width=0.3 \textwidth]{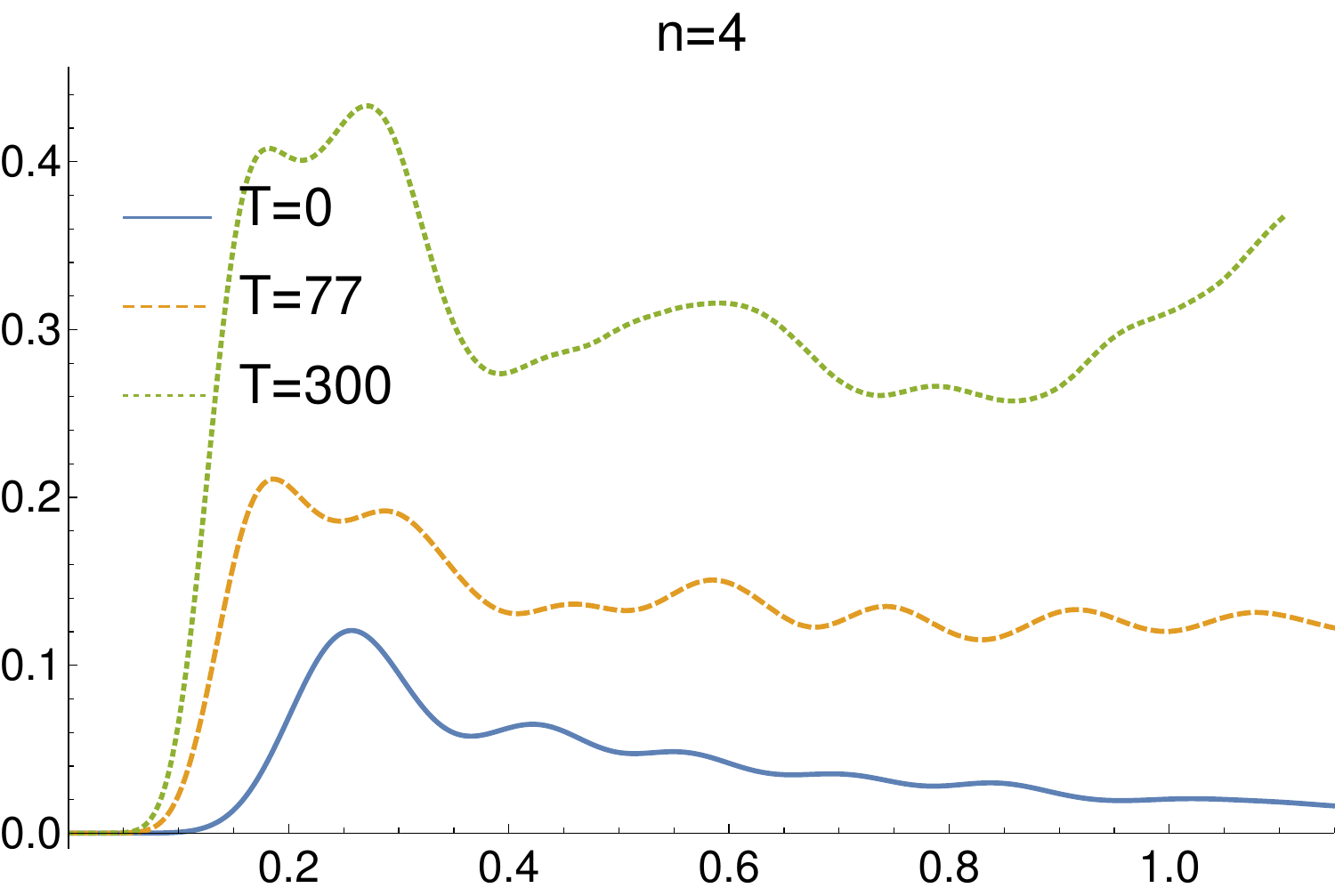}}
    \subfigure[]{\includegraphics[width=0.3 \textwidth]{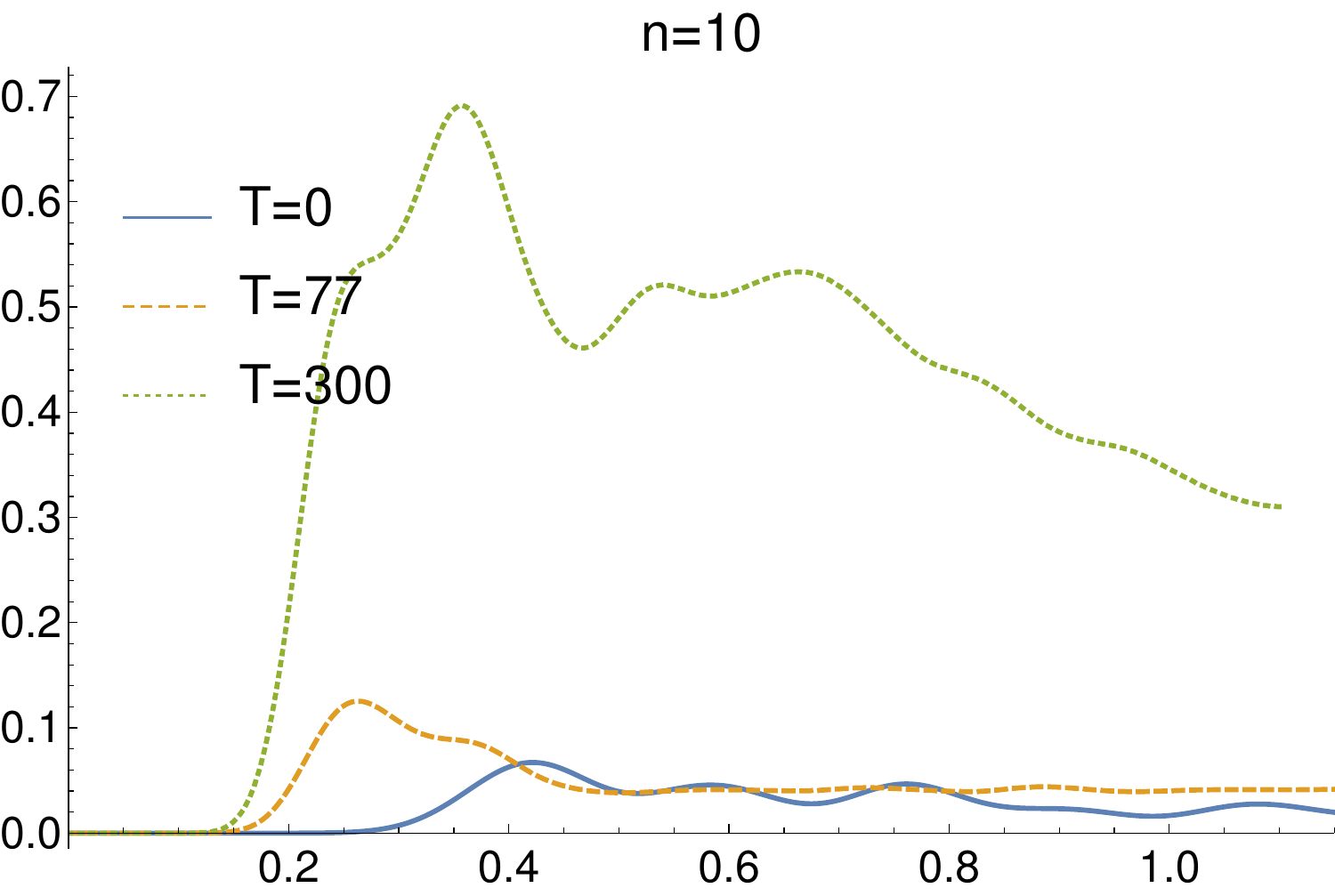}}
    \subfigure[]{\includegraphics[width=0.3 \textwidth]{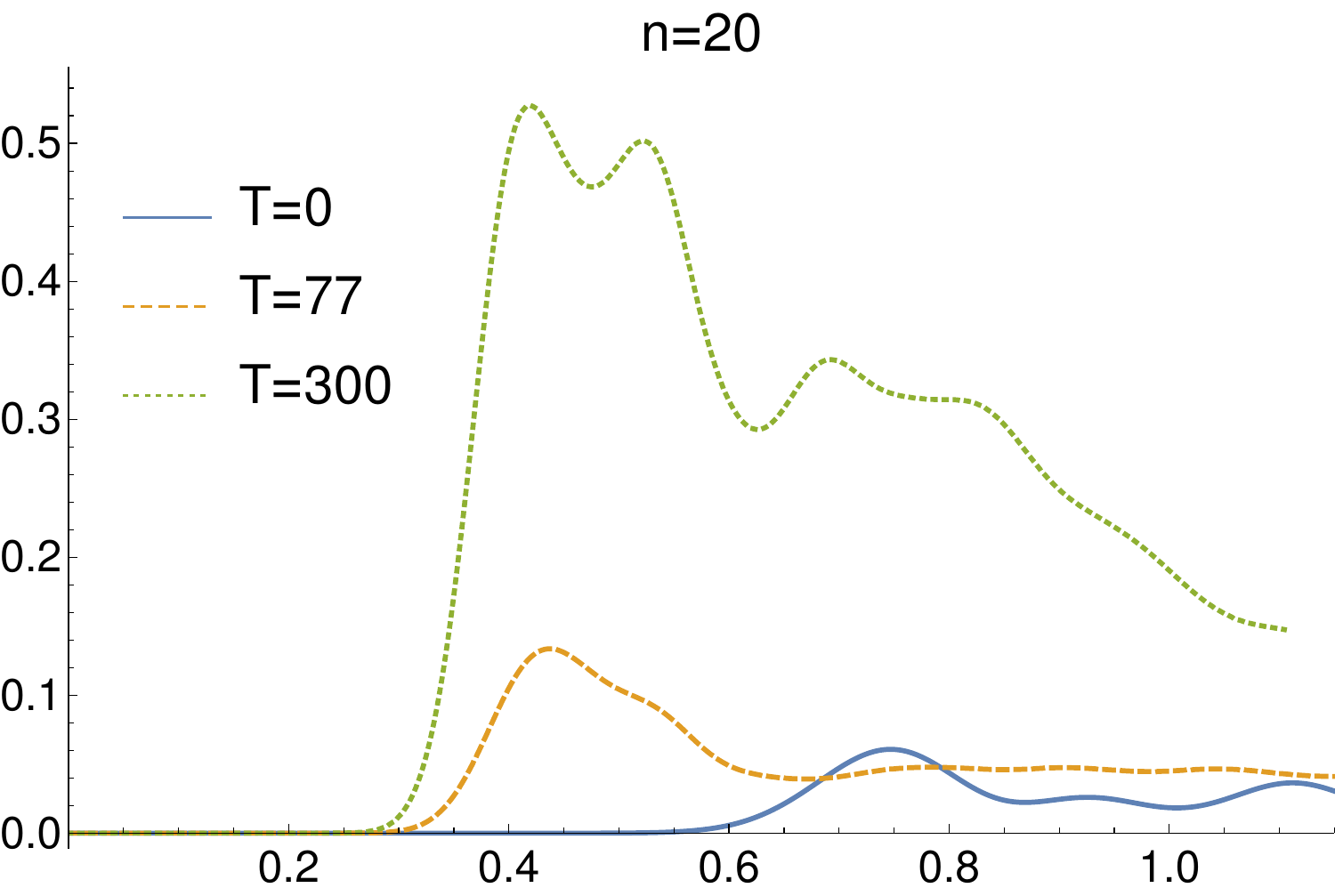}}
    \caption{\label{fig:occupationN} The average occupation number of the $n$-th site of the T-SD as a function of time and at different temperatures. (a) $n=0$, (b) $n=1$, (c) $n=4$, (d) $n=10$, (e) $n=20$. The plot of the same quantity for $n=30$ is shown in Fig.3(b) of the main text.}
\end{figure*}

\subsection{TEDOPA vs T-TEDOPA algorithmic complexity: details}\label{sec:vs}
Here, relying on the analyses in the previous Sections, we provide a more detailed description
{of the computational complexity scaling} of TEDOPA and T-TEDOPA. We will consider only the
complexity of the real-time evolution part and disregard the determination of the initial state
of the thermalized chain required by standard TEDOPA.

For {a spectral density $J(\omega)$ and inverse temperature $\beta$, we indicate by the chain
truncation $N$, local dimension $d$ and bond dimension $\chi$ that achieve converged results in the
simulation interval $[0,t_\text{max}]$ for standard TEDOPA, and by $N'$, $d'$, $\chi'$ the analogous}
parameters for T-TEDOPA. As proved in \cite{schollwock11}, the {computational} complexity of
TEDOPA is $O(N t_\text{max} (d^2 \chi)^3)$. The term $(d^2 \chi)^3$ is due to the computational cost of the
two-site update part of TEBD \cite{schollwock11}, which requires singular value decompositions (SVDs)
of $d^2 \chi$ matrices {(for simplicity we are assuming all the local and bond dimensions to be
constant along the chain)} and represents the real bottleneck of TEBD which absorbs, on the average,
about 90\% of the computational time. By adopting a randomized version of SVD (RSVD) \cite{tama15,tama18},
the complexity of TEDOPA can be {reduced} to $O(N t_\text{max} (d^2 \chi)^2)$. Even with {RSVD},
the simulations at high temperature are computationally highly demanding, since, $d$ needs to be taken
quite large at high $T$.


As mentioned in the main text, with T-TEDOPA the matrices that need to be handled are much smaller
for two reasons. First of  all the states to be represented are pure, so that MPS representation
suffices. {Moreover, typically, the local dimension $d'$ can be chosen significantly smaller
than $d$ at finite $T$. Local dimensions $d'=6,8,12$ for the for the monomer simulation at $T = 0,
77$ and $300K$ respectively,} for example, provided converged results, while the mere representation
of the chain thermal states would have required much larger local dimension. The availability of
tight estimates of the T-TEDOPA optimal local dimension would clearly be a most useful tool, and
{stringent mathematical results will be discussed elsewhere}.

In addition to this, {as discussed in the previous section, a further significant decrease in
the computational complexity can be reached via the use of a non-uniform local dimension $d'(n),
\ n=1,2,\ldots, N'$.  On the one hand, the occupation of the first sites in the chain, where localization of
the population injected from the system  occurs, is larger than the occupation in the remaining part
of the chain. Our numerical experience shows that with the choice $d'(n) = d'_{\text{max}} - n
(d'_{\text{max}}-2)/N'$ always leads to converged results. While the asymptotic scaling is left
invariant by this choice, the actual constants are much smaller than for the case of a uniform local
dimension $d'(n)=d'_{\text{max}}$.  This has a significant impact on the simulation time. In the original
TEDOPA scheme, an analogous fine tuning of the local dimension cannot be achieved with the same efficiency.
While the excitations that the system injects into the chain will still concentrate in the first
few sites of the chain, they distribute faster across the chain leading to significant occupation
numbers along the entire chain, with a maximum around the half of the chain (see Fig.2(a) of the main text as an example).
This reduces strongly the possibility to decrease the local dimension of the oscillators without detailed
knowledge about the system-environment dynamics.

On the other hand, while in TEDOPA the system-chain coupling $\kappa_0$ and the chain parameters $\omega_n,
\kappa_n \ n=1,2,\ldots$ are fixed, and only the initial state of the chain {was a function of
temperature, in T-TEDOPA all the $\omega_n, \kappa_n$} are temperature dependent. In particular, the
system-chain coupling is a monotonically increasing with temperature, $\kappa_{\beta,0} = ||J_\beta||_1 =
\sqrt{\int_{-\omega_c}^{\omega_c} d\omega J_\beta(\omega)}$ and $\kappa_{\beta,0} \stackrel{\beta
\to 0}{\longrightarrow} \sqrt{2} ||J||_1$. At high temperatures, therefore, the system-chain coupling
is larger for T-TEDOPA than for standard TEDOPA.
It is difficult to obtain a precise estimate for the required $\chi$ ($\chi'$) for T-TEDOPA and standard
TEDOPA and it is rather difficult to provide any general quantitative statement about the relation between
the bond dimensions in the two approaches. However, for all the numerical examples that we have considered
here, we always obtained T-TEDOPA converged results by setting at most $\chi'=1.4\chi$.}

Finally, due to the higher asymptotic value of the coupling between nearest-neighbour oscillators
in the {thermalized case, one has to take a longer chain, i.e. $N' > N$ (see the discussion in
Sect.\ref{sec:N}). In our numerical examples we have found that $N'\approx 2 N$ was always sufficient}
All in all, however, the use of a smaller local dimension, along with the possibility to decrease it
along the chain, amply compensate by itself the increased bond dimension and chain length.

We conclude this section with a technical, but relevant, remark. T-TEDOPA differs from TEDOPA only
in the computation of the chain coefficients. As such, it can be used with existing and optimized
TEDOPA codes. In TEDOPA, however, the workload related to the SVD decomposition and other matrix
operations on the
typically large (MPO) matrices was distributed over the available computing cores via multi-threaded
executions at an open-MP level (e.g. multi-threaded Intel Math-Kernel-Library \cite{mkl}). With the reduced dimension of the matrices
needed by T-TEDOPA, on the other side, such approach would not fully exploit the available computational
resources. This suggests to use the same computing cores at an openMP           \cite{openmp} or MPI \cite{mpi} level:
independent two-sites updates can be distributed over different threads or      processes running on multi-core architectures. A benchmark of our TEDOPA code with openMP parallelization showed a  linear speedup with the number of available cores w.r.t. single-core            executions. A TEDOPA code with MPI layer, distributing the workload over        different computational nodes, is currently under development.
\subsection{WSCP spectral density}
The spectral density $J_W(\omega)$ is the combination of a broad background at low frequency and
three Lorentzian peaks at high frequency. More specifically,
\begin{align}
    J_W(\omega) &= \sum_{k=1}^3 J_{LN,k}(\omega) + \sum_{m=1}^3 J_{L,m}(\omega), \\
    J_{LN,k}(\omega) & = \frac{S_k}{\sigma_k \sqrt{2 \pi}} \omega \exp \left \{-
        \frac{\left [\ln(\omega/\omega_k)\right ]^2}{2 \sigma_k^2} \right\},\\
        J_{L,m}(\omega) &= \frac{4 \gamma_m \Omega_m g_m (\Omega_m^2+\gamma_m^2)
        \omega}{\pi \llrrq{\gamma_m^2 +(\omega + \Omega_m)^2}\llrrq{ \gamma_m^2
        +(\omega -\Omega_m)^2}}.
    \label{eq:spectralFull}
\end{align}
The low frequency part (or backgound) is the
combination \cite{curutchet15} of three log-normal functions with
with $S_1=0.39, S_2=0.23, S_3=0.23$, $\sigma_1=0.4, \sigma_2=0.25, \sigma_3=0.2$, $\omega_1=26
\text{cm}^{-1}, \omega_2=51 \text{cm}^{-1},\omega_3= 85\text{cm}^{-1}$. The three Lorentzian peaks
have all the same width $\gamma_k = \gamma = 5 \text{cm}^{-1}$, and are centered in $\Omega_1 =181\text{cm}^{-1},
\Omega_2=221\text{cm}^{-1}, \Omega_3=240\text{cm}^{-1}$ and have weights $g_1 = 0.0173 , g_2 =0.0246, g_3 = 0.0182$.

\subsection{{Thermofield, T-SD and double-chain method}}
{In this section we discuss in more detail the relation {between the approach developed in this
work using a thermal spectral density and a pure state environment and the methods developed in
\cite{diosi98, yu04,blasone11,deVega15} which rely on the thermofield formalism. On the one hand,
this will allow us to show how to recover, at least in principle, the global state
of the original system at a generic time $t$. On the other hand, we will point out the differences,
in terms of simulation complexity, between T-TEDOPA and the numerical {thermofield} approach formulated in
\cite{deVega15}}.}

Consider the Hamiltonian
\begin{align}
    H_{SE} &= H_S + H_E + H_I\\
    H_E &= \int_0^{+\infty} d\omega \ \omega a_\omega^\dagger a_\omega;\\
    H_I  &= A_S \otimes \int_0^{+\infty} d\omega \ \sqrt{J(\omega)} X_\omega, \label{eq:suphi}
\end{align}
which is exactly the same as the one given in equations (1) and (2) of the main text, with the
{environment} part of the interaction Hamiltonian specialized to the position operator
$X_{\omega} = a_{\omega} + a_{\omega}^\dagger$, and $J(\omega)$ a spectral density. The initial
state is a product state $\rho_S \otimes \rho_E$, with each environmental mode in a thermal state
at inverse temperature $\beta$. The first step of the thermofield approach is a purification of
the state of the environment. This can be accomplished through the introduction of a set of new
modes with annihilation and creation operators $b_\omega, b_\omega^\dagger,\ \omega \geq 0$ satisfying
the standard bosonic commutation relations. The additional modes form an additional environment
$E'$ that {do not interact with either the system or} with $E$. {The thermofield approach
choses the initial pure state such that the reduced state of $E$ is a thermal state with inverse
temperature $\beta$.} The Hamiltonian for this extended system is $H' = H_{SE} - \int_0^\infty
d\omega\  \omega b_\omega^\dagger b_\omega$, {so that $E'$ interacts neither with the system $S$
nor with the environment $E$}. {Now, a Bogoliubov transformation,} combining the original modes
in $E$ and in $E'$ into new bosonic modes
\begin{align}
    c_\omega &= \cosh(\theta_\omega) a_\omega - \sinh(\theta_\omega) b_\omega^\dagger, \\
    d_\omega &= \cosh(\theta_\omega) b_\omega - \sinh(\theta_\omega) a_\omega^\dagger,
    \label{eq:therBogo}
\end{align}
with $\theta_\omega$ satisfying $\cosh(\theta_\omega) = \sqrt{1+n_\omega(\beta)}$,
{allows one to get a unitarily equivalent system
with Hamiltonian}
\begin{align}
    \tilde{H} &= H_S +  \int_0^{+\infty} d\omega \ \omega \llrr{c_\omega^\dagger c_\omega -
    d_\omega^\dagger d_\omega} + \nonumber \\
    &+A_S \otimes \int_0^{+\infty} d\omega \ \cosh(\theta_\omega) \sqrt{J(\omega)}
    \frac{c_\omega+c_\omega^\dagger}{\sqrt{2}} \nonumber\\
    &+A_S \otimes \int_0^{+\infty} d\omega \ \sinh(\theta_\omega) \sqrt{J(\omega)}
\frac{d_\omega+d_\omega^\dagger}{\sqrt{2}},
    \label{eq:hamBogo}
\end{align}
{while the initial thermal state is mapped to the vacuum state of the two newly defined modes}, i.e.
$\langle c_\omega^\dagger c_\omega \rangle = \langle d_\omega^\dagger d_\omega \rangle=0$. We
observe that \eref{eq:hamBogo} is exactly the Hamiltonian derived in the main text for the
extended environment, with a formal separation between the contribution of the positive and
negative frequency oscillators.
{Moreover, since the Bogoliubov transformations in Eq.\eqref{eq:therBogo}
guarantee a unitary equivalence, in principle one can even recover the global
system-environment state at a generic time $t$, including
the environmental state, as well as the system-environment correlations.
In fact, given the state evolved under the Hamiltonian in Eq.\eqref{eq:hamBogo}
at a generic time $t$,
the inverse of Eq.\eqref{eq:therBogo} would give the global $S-E-E'$ state evolved
under $H'$ and then, after tracing over $E'$, the $S-E$ state of the system and the original environment at
time $t$. While this might be generally a practically demanding
operation to perform, it shows that the thermofield
approach, as well as of course the T-TEDOPA, still includes {the complete} information,
not only about the open system {of interest}, but also about the original global system.}

Now, the construction proposed in
\cite{deVega15} relies more directly on the Hamiltonian in Eq.\eqref{eq:therBogo},
and {it can be viewed essentially as performing
the TEDOPA chain mapping into two environmental oscillator chains}. Since these are initially in a pure state, MPS can be
used as long as the system is itself in a pure state, with a substantial computational advantage.

However, the need of two chains for each environment can have a major impact on numerical
simulations, compared to the T-TEDOPA approach we formulated here.
Beside the larger number of oscillators that need to be simulated, the presence of two
chains for each system part interacting with a local environment modifies the tensor structure, and
drastically increases the computational cost of all the update operations that involve the system. 
{For this reason, it is convenient to replace the double chains with a single one, as T-TEDOPA allows us to do
by means of the overall transformation on the spectral density given by Eq.(6) in the
main text. Nevertheless, let us emphasize that in order to introduce one single chain
it is crucial that the operators appearing in the interaction Hamiltonian are self-adjoint, analogously to what happens for the mapping of a finite-temperature 
non-Markovian quantum state diffusion \cite{diosi98}. If the interaction operator is not self-adjoint, as
would be for example after applying the rotating wave approximation to the Hamiltonian in Eq.(13), the positive and negative frequencies of the transformed Hamiltonian have to be treated separately, as in \cite{deVega15}.}


Consider, for example, the dimeric structure shown in \Fref{fig:dimerMap}(a), where we have two
two-level systems interacting with each other and with two local environments that, for the sake of
simplicity and without loss of generality,  have the same spectral density $J(\omega)$. The
system resulting from double-chain and T-TEDOPA mappings are shown in \Fref{fig:dimerMap}(b)
and (c) respectively; the corresponding tensor structures are
show in \Fref{fig:dimerTens}. The polygons represent the tensor associated by the MPS description
to each TLS/oscillator; the degree of
the tensors is encoded in the number of edges of the polygons, the circles represent the bonds
between such tensors and the loose edges indicate the physical index of the tensor. The indices of
the tensor run over different ranges: the physical index range is $\{1,\ldots,d\}$ where $d$ the number of
internal states, the local dimension, of the system the tensor is associated to; the range of
the indices corresponding to the bonds is $\{1,\ldots,\chi\}$, with $\chi$, the bond dimension.

\begin{figure}[t]
    \includegraphics[width=1.\columnwidth]{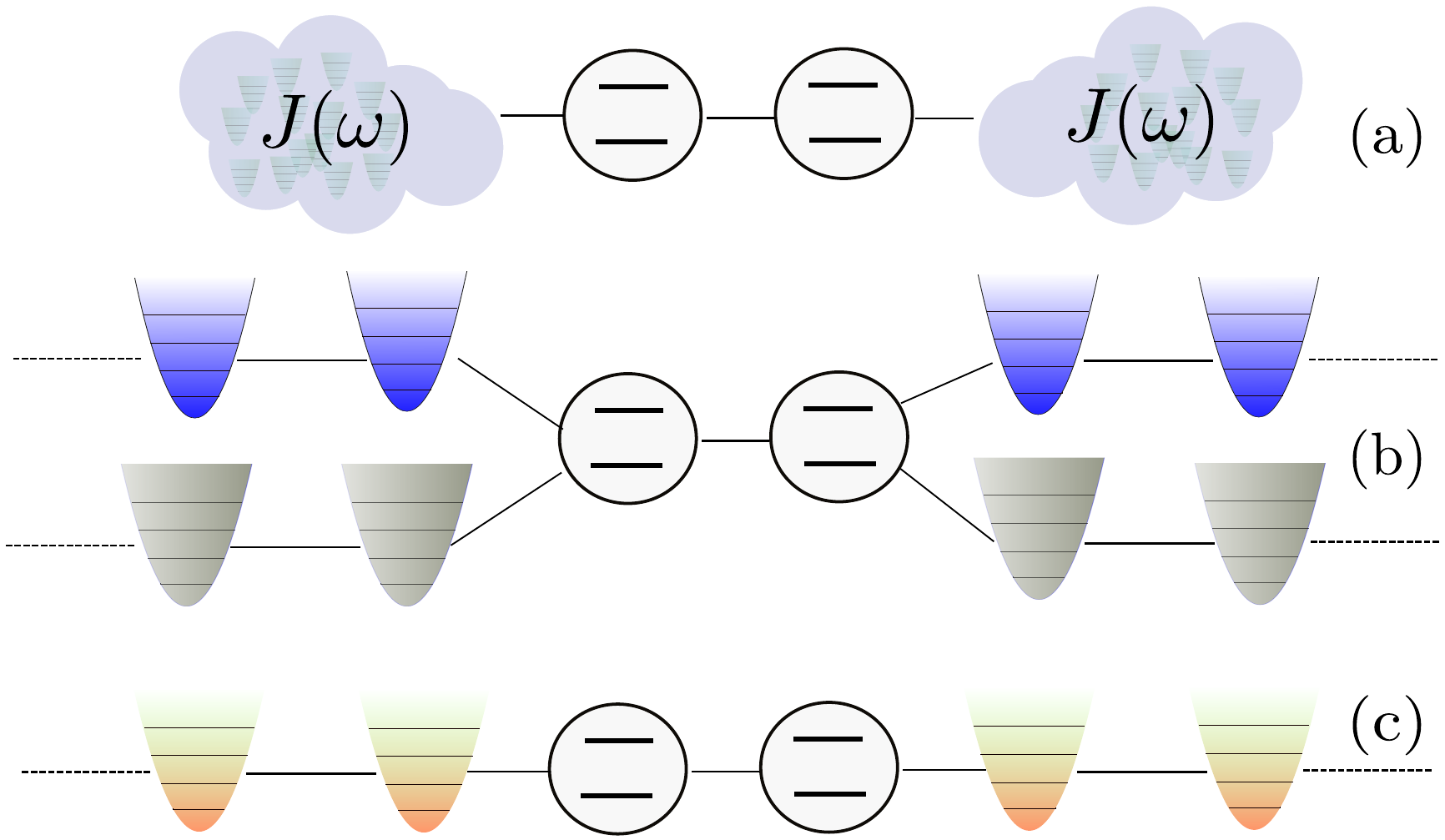}
    \caption{\label{fig:dimerMap}{Comparison with the chain mapping of \cite{deVega15}. (a) Dimeric system subject to local environmental interaction. (b) The resulting chain-mapping with the double-chain approach. (c) The resulting chain mapping with T-TEDOPA;
    {note that self-adjoint interaction operators are assumed, to halve the number of chains involved}.}}
\end{figure}

It is clear that the tensors associated to the two system sites by the
double-chain approach have rank 4 whereas with T-TEDOPA they have rank 3. A two-site update of nearest
neighbor tensors, say the $i$-th  and the $j$-th with rank $r_i$ and $r_j$, requires the
contraction of the two involved tensors into a single tensor. This operation produces a new tensor
of rank equal $r_i +r_j-2$, which is then reshaped into a $m \times n$ matrix, where $n$ and $m$
depend on the physical and bond dimensions of the tensors. An operator is then applied to this
matrix and the resulting matrix is suitably decomposed into two tensors. An essential step of
this procedure
is the already mentioned SVD of the $m\times n$ matrix, which, if $n\leq m$, has complexity $O(m
\cdot n^2)$. If we consider, for the sake of simplicity, the same dimension $\chi$  for all the
tensor network bonds, it turns out that the complexity of of the two site update of the tensor
corresponding to the dimers is $(2\chi) \cdot (2 \chi)^2 = 2^3 \cdot \chi^3$ for the T-SD
configuration, and $(2\chi^2) \cdot (2 \chi^2)^2 = 2^3 \cdot \chi^6$ for the double-chain one.

It is clear that, even though the increased complexity of the local updates concerns only those
updates that involve the TLSs, T-TEDOPA provides a dramatic improvement of the overall simulation time.
Just to give an example, if the bond dimension is set to $\chi=50$, as it typically needs to be
in the case of strong TLS-TLS or TLS-environment coupling, a single two-site update of the dimer
would require  $10^6$ operations with T-TEDOPA and $10^{11}$ operations with the double-chain method.

\begin{figure}[t]
    \vspace{10pt}
    \includegraphics[width=1.\columnwidth]{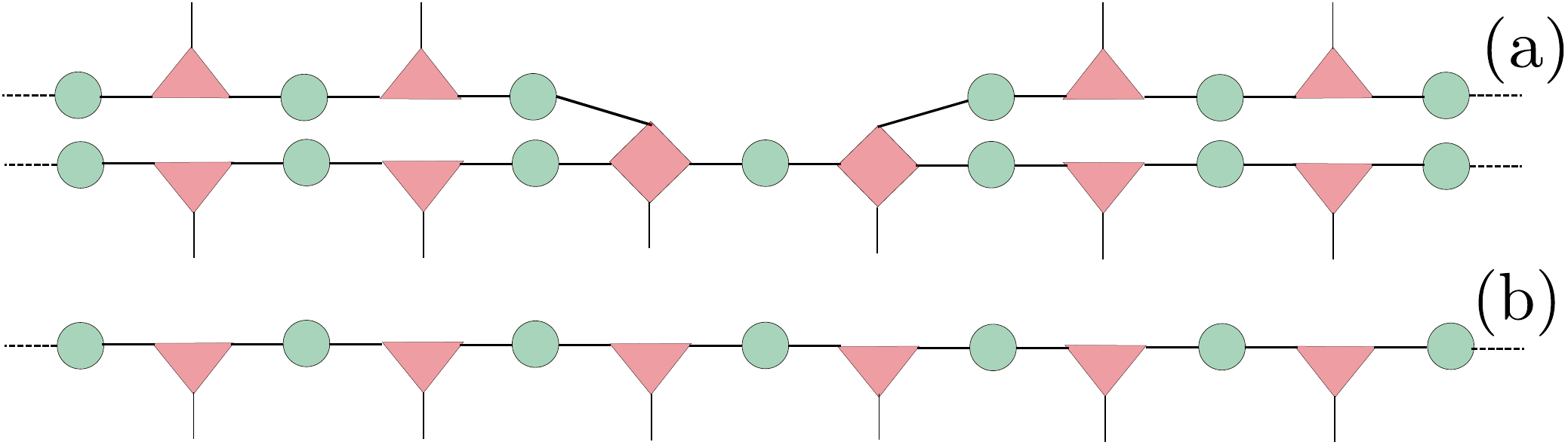}
    \caption{\label{fig:dimerTens} The tensor structure corresponding to (a) the double-chain approach, (b) T-TEDOPA, with graphical conventions as explained in the text.}
\end{figure}

\subsection{Hierarchical equations of motion}

{In Fig.2(b) in the main manuscript, we provide HEOM results for a comparison with T-TEDOPA for structured $J_W(\omega)$ and less-structured $J_W'(\omega)$ spectral densities at $300\,{\rm K}$. The parameters of the HEOM simulations are determined by the multi-exponential fitting of the two-time correlation function \cite{kubo89,tanimura06,lim18}
\begin{align}
    S(t) &= \int_0^{+\infty} d\omega\  J(\omega) \llrrq{e^{-i\omega t} (1+n_\omega(\beta))+ e^{i \omega t}
n_\omega(\beta)}
    \nonumber\\
    &\approx \sum_{k=1}^{X/2}(A_k e^{(i\Omega_k-\Gamma_k)t}+B_k e^{(-i\Omega_k-\Gamma_k)t}),
\end{align}
where $A_k$ and $B_k$ are independent, complex-valued amplitudes, while $\Omega_k$ and $\Gamma_k$ are real-valued frequencies and damping rates of the exponential terms. 
\begin{figure}[t]
    \vspace{10pt}
    \subfigure[]{\includegraphics[width=.49 \columnwidth]{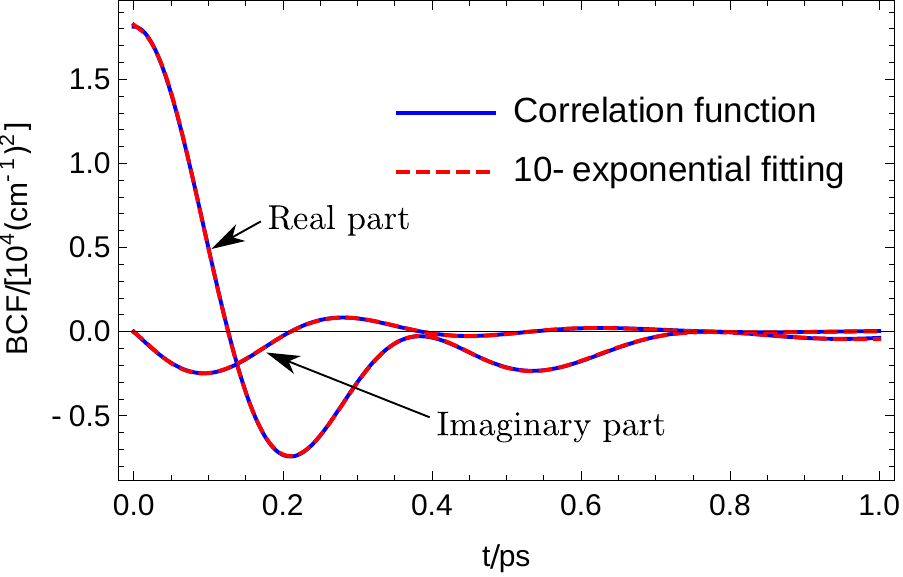}}\subfigure[]{\includegraphics[width=0.236 \textwidth]{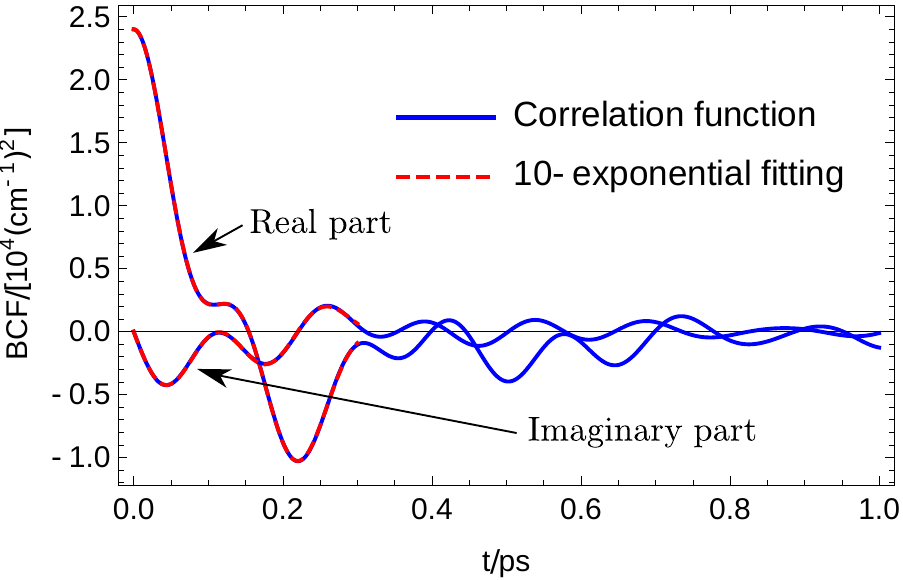}}
    \caption{\label{fig:HEOM_BCF} Two-time correlation functions at $T=300\,{\rm K}$. (a) For the less-structured $J_W'(\omega)$, the correlation function is fitted by 10 exponentials up to $1\,{\rm ps}$. (b) For the structured $J_W(\omega)$, including additional three Lorentzian peaks, the correlation function becomes more oscillatory than (a) with additional multiple frequency components. To maintain the fitting quality with the same number of exponentials, the fitting is performed up to $0.3\,{\rm ps}$, which determines the simulation time of HEOM (see Fig.2(b) in the main manuscript).}
\end{figure}
We note that the number of exponentials is one of the dominant factors determining the HEOM simulation cost. In \Fref{fig:HEOM_BCF}(a), the two-time correlation function of the less-structured spectral density $J_W'(\omega)$ is shown, which is fitted by a sum of 10 exponentials ($X=10$). The difference between target $S(t)$ and fitting function is minimized in such a way that the difference is more than two orders of magnitude smaller than the amplitude of the target function $S(t)$ up to $1\,{\rm ps}$. On the other hand, \Fref{fig:HEOM_BCF}(b) shows the case of the structured spectral density $J_W(\omega)$ with additional three Lorentzian peaks included, where the fitting is performed only up to $0.3\,{\rm ps}$, so that the target function $S(t)$ can be fitted by the same number of exponentials ($X=10$). This means that the optimized fitting parameters enable one to perform reliable HEOM simulations only up to $0.3\,{\rm ps}$. We note that at longer times, the correlation function starts to show additional frequency components, for instance when Fourier-transformed, not due to the ringing artifacts induced by a finite time window, but due to the presence of multiple modes in $J_W(\omega)$. This requires one to introduce additional exponential terms to maintain the fitting quality, which increases the HEOM simulation cost significantly. For both spectral densities, the number of auxiliary operators is increased until the system dynamics shows convergence, which is achieved at tier 14.}

%

%

\end{document}